\documentclass[aps,prl,twocolumn,float]{revtex4}
\usepackage{amsmath,bm,epsfig}
\usepackage{amsmath}
\usepackage{pstricks}
\usepackage{pst-node}
\usepackage[ansinew]{inputenc}
\usepackage{amssymb,amsmath}
\usepackage{amsfonts}
\usepackage{epsfig}
\usepackage[english]{babel}
\usepackage{graphics}

\def\a{\alpha}

\def\g{\gamma}
\def\d{\delta}
\def\o{\omega}

\def\e{\varepsilon}

\font\Sets=msbm10

\def\Real {\hbox{\Sets R}}

\def\Natural {\hbox{\Sets N}}
\def\Rational {\hbox{\Sets Q}}

\def\be{\begin{equation}}
\def\ba{\begin{array}}
\def\ee{\end{equation}}
\def\ea{\end{array}}
\def\bea {\begin{eqnarray}}
\def\eea {\end{eqnarray}}
\def\bean{\begin{eqnarray*}}
\def\eean{\end{eqnarray*}}

\def\const {\mathop{\rm const}\nolimits}

\def\RA {\ \Rightarrow\ }

\begin{document}

\title{Exact and quasi-resonances in discrete water-wave turbulence}

\author{Elena Kartashova}
\email{lena@risc.uni-linz.ac.at}
\affiliation{RISC, J.Kepler
University, Linz 4040, Austria }

\begin{abstract}
The structure of discrete resonances in water-wave turbulence is
studied. It is shown that the number of exact 4-wave resonances is
huge (hundreds million) even in comparatively small spectral domain
when both scale and angle energy transport is taken into account. It
is also shown that angle transport can contribute inexplicitly to
scale transport. Restrictions for quasi-resonances to start are
written out. The general approach can be applied directly to
mesoscopic systems encountered in condensed matter (quantum dots),
medical physics, etc.

{\it Pacs: 47.35.Bb, 89.75.Kd}
\end{abstract}


\maketitle

{\bf 1. Introduction.} Statistical wave turbulence theory deals with
the fields of dispersively interacting waves. Examples of these wave
systems can be found in oceans, atmospheres, plasma, etc.
Interactions between waves are similar to interactions between
particles and can be described by kinetic equations (KEs) analogous
to KE known in quantum mechanics since 1930th. Wave KE is in fact
one limiting case of the quantum Bose-Einstein equation while the
Boltzman kinetic equation is its other limit. First wave KE for
surface gravity waves has been presented in \cite{has} while general
method of construction KEs for many other types of waves can be
found in \cite{lvov}. One of the major achievements of the
statistical approach is establishing of the fact that power energy
spectra (similar to famous Kolmogorov 5/3 law) are exact stationary
solutions of kinetic equations \cite{zak2}. The limitations of
statistical turbulence theory are due to the fact that it does not
describe spatial unevenness of turbulence, i.e. organized structures
extending over many scales, like boulders in a waterfall, remain
unexplained. Appearance of these structures is attributed to
 mesoscopic regimes which are at the frontier between classical (single
waves/particles) and  statistical (infinite number of
waves/particles) description of  physical systems. Mesoscopic
(effectively zero-dimensioned) systems is very popular topic in
various areas of modern physics - from wave turbulence to condensed
matter (quantum dots, \cite{dots}) to medical physics. For instance,
in \cite{med1} dynamics of blood flow in humans is studied,
cardiovascular system is described by a few coupled oscillators and
synchronization conditions are investigated. Synchronization or
resonance conditions have the same general form for wave and quantum
systems (see, for instance, \cite{spohn} for 4-photon processes);
and have to be studied in integers. Just for simplicity of
presentation we prefer to stay with wave terminology while all
examples in this Letter are taken from wave turbulent systems.

Resonance conditions have the form
 \be \label{res} \o_1 \pm \o_2 \pm ... \pm \o_s=0, \quad
\vec{k}_1 \pm \vec{k}_2 \pm .... \pm \vec{k}_s=0 \ee where $ \
\o_i=\o(\vec{k}_i), \ \ s< \infty,\ $ with $\ \vec{k}\ $ and
$\o_i=\o(\vec{k}_i)$ being correspondingly wave vector and
dispersion function. Specific features of these systems described by
Fourier harmonics with integer mode numbers were first presented in
\cite{PRL} (we call them further discrete wave systems, DWS).
 A counter part to
kinetic equation in DWS is a set of
 few {\it independent} dynamical systems of ODEs on the amplitudes of
 interacting waves. Mathematical theory of DWS was
developed in \cite{AMS} with general understanding that discrete
effects are only important in some bounded part of spectral space,
$\ |\vec{k}| < k_0, \ $ with some small finite $\ k_0 \ $, while the
case $\ |\vec{k}| >> k_0 \ $ is covered by kinetic equations and
power-law energy spectra. This general opinion was broken recently
as result of numerical simulations with Euler equations for
capillary waves \cite{zak3} and for surface gravity waves
\cite{zak4} where discrete clusters of waves were observed
simultaneously with statistical regime. Moreover, experimental
results \cite{DLN06} show that discrete effects are major and
statistical wave turbulence predictions are never achieved: with
increasing wave intensity the nonlinearity becomes strong before the
system loses sensitivity to the $\ \vec{k}$-space discreteness.

{\bf 2. Discrete wave systems.} In \cite{lam} a model is presented
which explains the appearance of discrete wave clusters in the large
spectral domains with $\ |\vec{k}| >> k_0. \ $ It shows that energy
power spectra
 are valid {\it not} in all
spectral domain with big $\ \vec{k}\ $ but have "holes" all over the
spectrum, in some integer $\ \vec{k},\ $ which describe discrete
dynamics of mesoscopic regimes. This understanding put ahead a novel
computational problem - computing integer solutions of (\ref{res})
in the large spectral domains. Indeed, (\ref{res}) turns into \vskip
-0.5cm \be \label{4grav}
\sqrt{k_1}+\sqrt{k_2}=\sqrt{k_3}+\sqrt{k_4},\quad
\vec{k}_1+\vec{k}_2=\vec{k}_3+\vec{k}_4 , \ee
 for
 4-wave interactions of 2{\bf D}-gravity water waves,
  where $\vec{k}_i=(m_i,n_i),
\ \forall i=1,2,3,4$ and $k_i=|\vec{k}_i|=\sqrt{m_i^2+n_i^2}$. This
means that in a domain, say $\ |m|,|n| \le D \sim 1000\ $, direct
approach leads to necessity to perform extensive (computational
complexity $\ D^8$) computations with integers of the order of
$10^{12}$. The full search for multivariate problems in integers
consumes exponentially more time with each variable and size of the
domain to be explored. The use of special form of resonance
conditions allowed us to develop fast generic algorithms
\cite{comp_all} for solving systems of the form (\ref{res}) in
integers. The main idea of the algorithms for irrational dispersion
function is that if vectors $\ \vec{k}_i\ $ construct an integer
solution of (\ref{res}), then at least for some $i_0,j_0$ the ratios
$\o_{i_0}/\o_{j_0}$ have to be  rational numbers. Some other
number-theoretical considerations were used in case of rational
dispersion functions.  In
 particular,
all integer solutions of (\ref{4grav})
 in domain $\ |m|,|n| \le
1000\ $ were found in a few minutes at a Pentium-3 (cf.: direct
search for them in smaller domain $\ |m|, |n|\le 128$ took 3 days
with Pentium-4 \cite{naz1}).

 Using our programs  we studied how the structure of discrete resonances depends
on the form of dispersion function and on the chosen $s$. The
interesting fact is that characteristic structure is the same for
different dispersion functions (examples with $\o=m/n(n+1), \
\o=1/\sqrt{m^2+n^2}, \ \o=m/(n^2+m^2+1)$ and others were studied) if
$s=3$. Most waves do not take part in resonances and interacting
waves form small independent clusters, with no energy flow between
clusters due to exact resonances.
 Our conclusion about wave systems with
3-wave interactions is therefore that exact resonances are rare and
quasi-resonances, i.e. those satisfying  \be \label{quasi}
 \o_1 \pm \o_2 \pm ... \pm \o_s=\Omega>0, \quad \vec{k}_1 \pm
\vec{k}_2 \pm .... \pm \vec{k}_s=0  \ee with $\ s=3\ $ can be of
importance for some applications. The situation changes
substantially in the case $\ s\ge 4 \ $  which is illustrated below
with surface gravity waves taken as our main example.

\begin{figure}[h]
\begin{center}\vskip -0.2cm
\includegraphics[width=6cm,height=3.5cm]{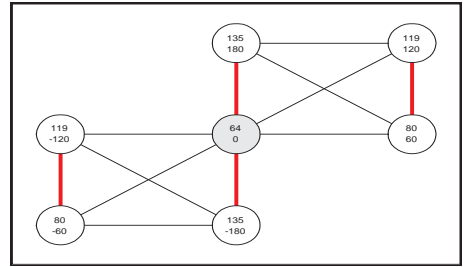}
\end{center} \vskip -0.6cm
\caption{Wave (64,0) takes part in 2 scale-resonances. The upper
number in the circle is $m$ and the lower - $n$, and red thick lines
drawn between vectors on the same side of the Eqs.(\ref{4grav}).
\label{f:Mul200A}}
\end{figure}

\begin{figure}[h]
\begin{center}\vskip -0.7cm
\includegraphics[width=7cm,height=3.7cm]{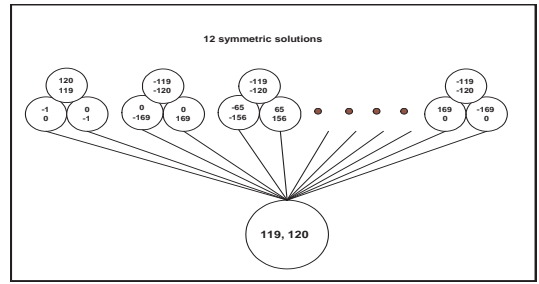}
\end{center} \vskip -0.6cm
\caption{ Wave (119,120) takes part in 12 angle-resonances.
\label{f:Mul200B}}
\end{figure}

{\bf 3. Exact resonances.} The major difference between 3- and
4-wave interactions can be briefly formulated as follows. Any 3-wave
resonance generates new wave lengths and therefore takes part in
energy transfer over scales. In a system with 4-wave resonances two
mechanisms of energy transport are possible: 1) over scales, if at
least one new wave length is generated, and 2) over angles, if no
new wave lengths are generated. Examples of
 these two types of solutions for (\ref{4grav}) are
$(-80,-76)(980,931)\RA(180,171)(720,684)$ and
$(-1,4)(2,-5)\RA(-4,1)(5,-2)$, we call these solutions further
scale- and angle-resonances correspondingly. Two mechanisms of
energy transport provide substantially richer structure of
resonances and the number of exact resonances grows enormously when
compared to 3-wave resonance system. Say, for dispersion
$\o=1/\sqrt{m^2+n^2}$ and 3-wave interactions there exist only 28156
exact resonances with $|m|,|n| \le 1000$ while for dispersion
$\o=(m^2+n^2)^{1/4}$ and 4-wave interactions in the same domain the
overall number of exact resonances is about 600 million. However,
major part of these resonances are angle-resonances. In the domain
$\ |m|,|n| \le 1000 \ $ we have found only 3945 scale-resonances,
i.e. transport over the scales is similar to the case of 3-wave
interactions. Some isolated quartets do exist, also among
angle-resonances, for instance $(-1,-1)(1,1)\RA(-1,1)(1,-1)$ but
they are rather rare. Very important fact is that angle and scale
energy transport are not independent in following sense. One wave,
say (64,0), takes part in 2 scale-resonances one of them being
$(64,0)(135,180)\RA(80,60)(119,120)$ and the wave (119,120) takes
part then in 12 angle-resonances, and further on (see
Figs.\ref{f:Mul200A},\ref{f:Mul200B}).

It is important to understand that for $\ s>4 \ $ 1) no new type of
resonances appear, and 2) existence of angle-resonances depends on
the placing of signs in the first equation of (\ref{res}): for
instance, if it has the form $\o_1 = \o_2 + ... + \o_s,$ no
angle-resonances are possible.

The natural question now is whether quasi-resonances are in fact of
importance in a wave system possessing such enormous number of exact
resonances.

{\bf 4. Quasi-resonances.} From now on we are interested in discrete
quasi-resonances which are integer solutions of (\ref{quasi}) with
some non-zero resonance width $\ \Omega. \ $ Notice that  $\ \omega
: \mathbb{Z} \times \mathbb{Z} \rightarrow \mathbb{Q}(q)\ $,  where
$\ q\ $ is an algebraic number of degree 4 and the field $\
\mathbb{Q}(q)\ $ denotes corresponding algebraic expansion of $\
\mathbb{Q}\ $. To estimate a linear combination of different $\o_i$
over $\ \mathbb{Q}\ $ we use generalization of the Thue-Siegel-Roth
theorem \cite{tue}: {\it If the algebraic numbers $\ \alpha_1,
\alpha_2, ...,\alpha_s\ $ are linearly independent with 1 over $\
\mathbb{Q}\ $,
 then for any $\ \e>0\ $ we have
$
 |p_1 \alpha_1 + p_2 \alpha_2 + ... +p_s\alpha_s - p|> c
 p^{-s-\e}
$ for all $\ p, p_1, p_2, ..., p_s \in \mathbb{Z}\ $ with $\ p =
\max_i | p_i |.\ $  The non-zero constant $\ c\ $ has to be
constructed for every specific set of algebraic numbers separately.}
For $\ \o=(m^2+n^2)^{1/4}\ $ and four-term combination this
statement means in particular that $\ |\o_1\pm\o_2\pm\o_3\pm\o_4|>1\
$ if at least one of $\ \o_i\ $ is not a rational number. As it was
pointed out in \cite{comp_all} all integer solutions of
(\ref{4grav}) have one of two forms: I) $\ \o_i=\g_iq^{1/4},\
\forall i=1,2,3,4\ $ or II) $\ \o_1=\o_3=\g_1q_1^{1/4}, \
\o_2=\o_4=\g_2q_2^{1/4}, \ q_1\neq \ q_2. \ $ Here $\ \g_i,q, q_1,
q_2\ $ are some integers and $\ q, q_1, q_2\ $ have form $\
p_1^{e_1} p_2^{e_2} ... p_n^{e_n}, \ $ with all different primes $\
p_1, ... p_n\ $ while the powers $\ e_1, ...e_n \in \Natural \cup
\{0\}\ $ are integers all smaller than $4$. It follows that for all
wave vectors but those of the form I) with $\ q=1,\ $ there exists
{\bf a global low boundary} for resonance width, $\ \Omega>1, \ $
necessary to start quasi-resonances. In the spectral domain $\
|m|,|n| \le 1000\ $ only 136 scale-resonances do not have global low
boundary. But even for them local low boundary exists - defined  by
the spectral domain  $\ T=\{(m,n): 0 < |m|,|n| \le D < \infty\}\ $
chosen for numerical simulations. Indeed, let us define $\  \Omega_D
=\min_p \Omega_p,\ $ where 
$
 \Omega_p =|\o(\vec k^{p}_1) \pm \o(\vec k^{p}_2) \pm ... \pm \o(\vec k^{p}_4)|,\ \
 \vec k^{p}_j = (m^{p}_j, n^{p}_j) \in T,$
for $\forall j=1,2,3,4, $  $\ \o(\vec k^{p}_1) \pm \o(\vec k^{p}_2)
\pm ... \pm \o(\vec k^{p}_4) \neq 0 \quad \forall p,\ $ and index $\
p\ $ runs over all wave vectors in $\ T\ $, i.e. $\ p\le 4D^2\ $. So
defined $\ \Omega_p\ $ obviously is a non-zero number as a minimum
of finite number of non-zero numbers and $\ \Omega_D\ $ is minimal
resonance width which allows discrete quasi-resonances to start, for
chosen $\ D.$

Step of numerical schema  $\ 0<\d\ll 1 \ $ is another parameter
important for understanding quasi-resonant regimes, and
inter-relation
 between $\ \d, \ $ $\ \Omega_D \ $ and $\ \Omega \ $ describes them
 all. For instance, if $\ \Omega_D>\Omega, \ $ any chosen $\ \d_0> \Omega_D\ $ will allow some number of
quasi-resonances, say $\ N_{(\d_o> \Omega_D)}\ $, and for any $\
\d>\d_0 \ \RA \ N_{(\d> \Omega_D)} \ge N_{(\d_o > \Omega_D)}.\ $ On
the other side, if $\ \d\ $ is decreasing to $\ \Omega_D\ $ from
above, $\ \d \rightarrow \Omega_D+0, \ $ the number of
quasi-resonances reaches some constant level $\ N_{min}\ $, $\
N_{min}=N_{(\d= \Omega_D)}\ $. If $\ \d \ $ is increasing to $\
\Omega_D\ $ from below, $\ \d \rightarrow \Omega_D-0, \ $ the number
of quasi-resonances is $\ N_{min}=\const. \ $

This fact has been first discovered in the numerical simulations
\cite{Tan2004}, both for capillary and surface gravity waves
(maximal spectral domain studied was $\ 0< m \le 2047, \ 0< n\le
1023\ $ and only scale-resonances were regarded). In case of gravity
surface waves, increasing of $\ \Omega \ $ from $10^{-10}$ to
$10^{-5}$ does not changes number $\ N\ $ of quasi-resonances while
increasing of $\ \Omega \ $ from $\ 10^{-4}\ $ to $\ 10^{-3}\ $
yields increasing $\ N \rightarrow 10N.\ $ It turned out that
limiting level $\ N_{min}\ $ is 0 for capillary  and and $\ \sim
10^5\ $ for surface waves, for the same discretization by a
rectangular mesh.  It was attributed to the following fact:
quasi-resonances are formed in some vicinity of exact resonances
which do exist in the case of gravity waves and are absent in the
case of capillary waves.

\begin{figure}[h]
\begin{center}\vskip -0.2cm
\includegraphics[width=6.8cm,height=2.8cm]{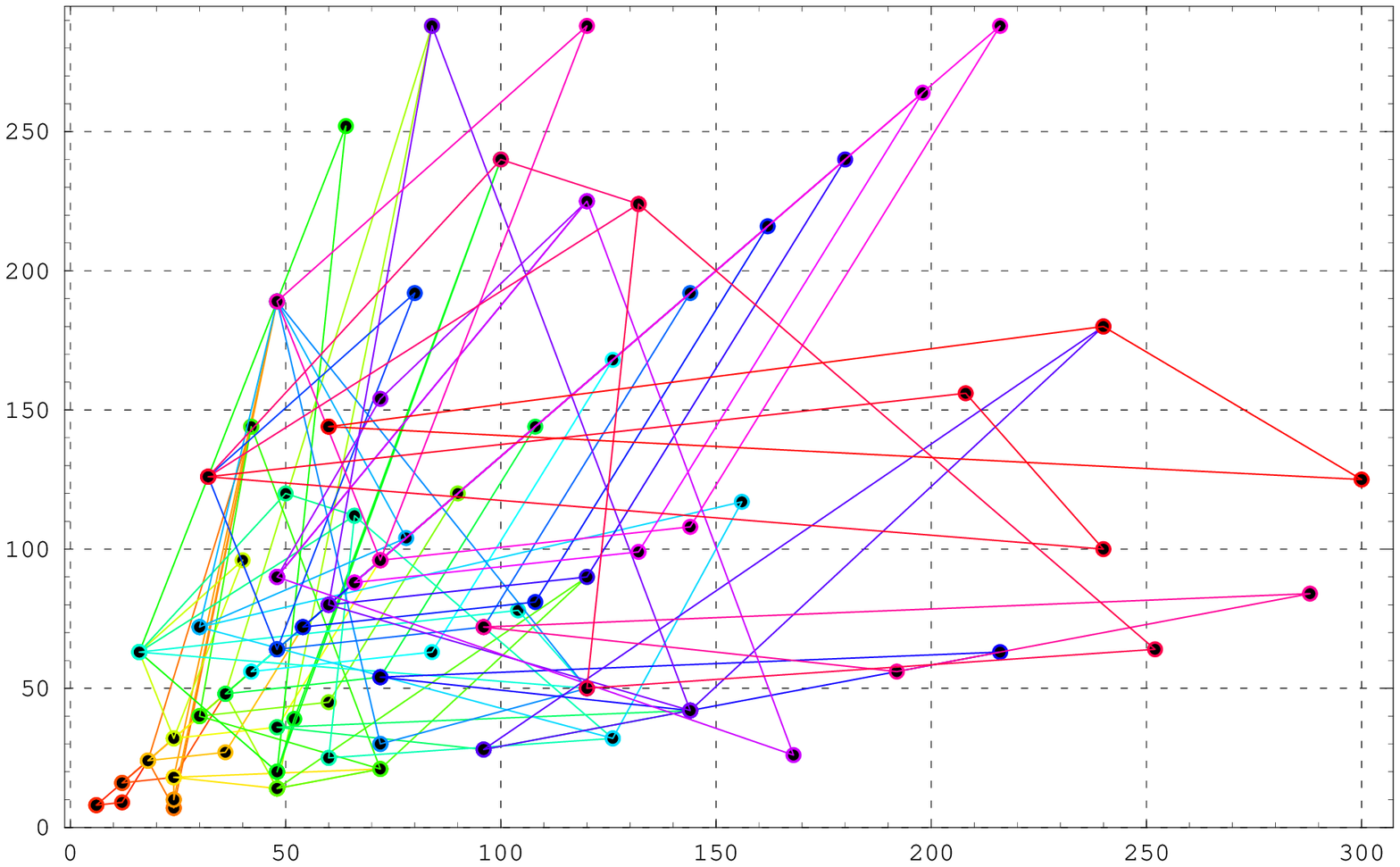}
\includegraphics[width=6cm,height=2.6cm]{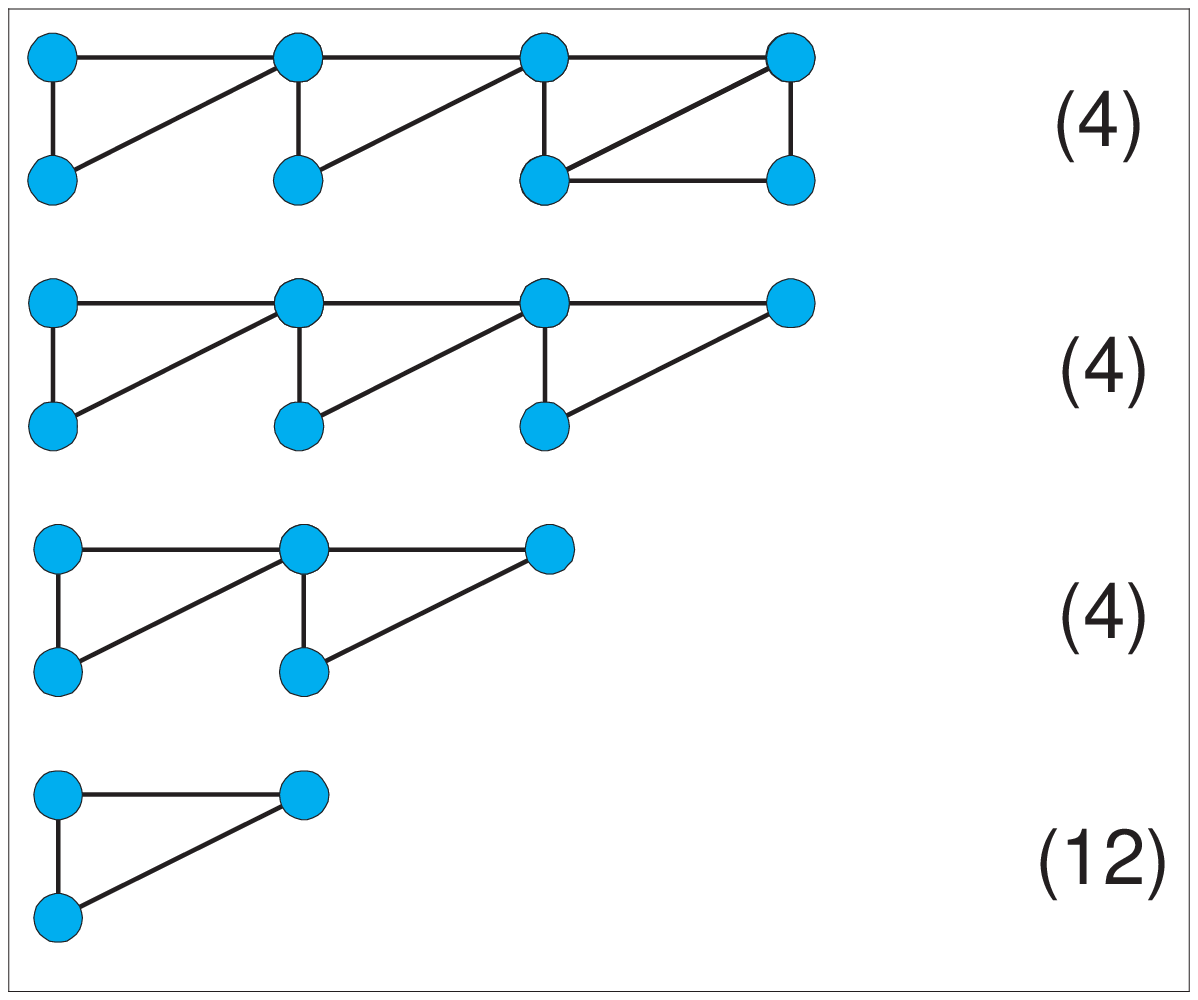}
\end{center} \vskip -0.6cm
\caption{\label{f:str1} {\bf Upper panel:} Geometrical structure,
$D=300$. {\bf Lower panel:} Topological structure, $D=50$. The
number in brackets shows how many times corresponding cluster
appears in the chosen spectral domain.}
\end{figure}

{\bf 5. Topological structure of resonances.} Graphical presentation
 of  discrete 2{\bf D}-wave clusters suggested in \cite{AMS} is to regard each
2D-vector as a node of integer lattice in spectral space and connect
those nodes which construct one solution (triad, quartet, etc.) We
demonstrate it in Fig.\ref{f:str1} (upper panel) taking for
simplicity 3-wave interactions with $\o=1/\sqrt{m^2+n^2}$ (ocean
planetary waves). Obviously, geometrical structure is too nebulous
to be useful. On the other hand, topological structure in
Fig.\ref{f:str1} (lower panel) is quite clear and gives us immediate
information about the form dynamical equations covering behavior of
each cluster. The number in brackets shows how many times
corresponding cluster appears in the chosen spectral domain. All
similar clusters are covered by similar systems of ODEs (written for
simplicity for real-valued amplitudes): $\ \dot{A}_1= \a_1 A_2A_3, \
 \dot{A}_2= \a_2 A_1A_3, \
 \dot{A}_3= \a_3 A_1A_2, \
$ in the case of a "triangle" group, two coupled systems of this
form in the case of "butterfly" group and so on.
 A 3-wave system has been chosen here as an
illustrative example for its simplicity. They have only one type of
vertices in their graphical presentations: nodes, corresponding to
exact resonances. Some graph-theoretical considerations allow to
construct isomorphism between topological elements of the solution
set of (\ref{res}) and corresponding dynamical systems, for the case
$\ s=3. \ $ This yields a constructive method to generate all
different dynamical systems in a given spectral domain. For
instance, all graphs shown in Fig.\ref{f:str1}, lower panel, are
described by 4 different dynamical systems; for all isomorphic
graphs, corresponding dynamical system has the same form though
coupling coefficients $\ \a_i \ $ have different magnitudes, of
course.
 Some programs are written
in MATHEMATICA which allow for a few specific examples to a) find
all integer solutions of (\ref{res}), b) generate their geometrical
and topological structure, c) write out explicitly all corresponding
dynamical systems. Only small number of these systems are known to
be solvable analytically (mostly those corresponding to clusters of
3 to 5 waves only) while larger systems should be solved
numerically, of course. Knowledge of specific form of a dynamical
system allows in many cases to write out some conservation laws and
thus simplify substantially further numerical simulations.

In case of $\ s$-wave interactions with $\ s\ge 4\ $ construction of
a corresponding graph must be substantially refined: a graph with 3
different types of vertices should be constructed, corresponding to
the waves taking part in 1) angle-resonances, 2) scale-resonances,
3) both types of resonances,
 in order to provide simultaneous isomorphism of graphs and
dynamical systems. This work is under the way.

Knowledge of the resonances structure might contribute to short-term
forecast of wave field evolution,
 for in direct
numerical simulations \cite{Tan2004},\cite{T07} discrete resonances
were observable not at the time scale $O(t/\e^4)$ of kinetic theory
($\e$ is steepness of wave field) but {\it at linear time scale}
$t.$

{\bf 6. Conclusions.}

{\bf 1.}  3-wave systems can possess only scale-resonances which are
rare, in this case quasi-resonances might be of importance in energy
transport;

{\bf 2.} $s$-wave systems with $\ s\ge 4,\ $ depending on the sign
setting in (\ref{res}), may also posses angle-resonances which
contribute inexplicitly into energy transport. In systems like
(\ref{4grav}) where angle-resonances are allowed, there exist
hundreds million of exact resonances in a comparatively small
spectral domain $\ |m|,|n| \le 1000.\ $

{\bf 3.} For some polynomial irrational dispersion function, global
low boundary $\ \Omega\ $ for quasi-resonances to start can be
computed which 1) does not depend on the chosen spectral domain, and
2) is valid for the most part of exact resonances in a system. If
dispersion is a rational function, only local low boundary $\
\Omega_D \ $ exists (it follows from the fact that $\ \Rational \ $
is dense everywhere in $\ \Real $).

{\bf 4.} Any interpretation of  results of numerical simulations
with dispersive wave systems has to take into account interplay
between $\ \d, \ $ $\ \Omega_D \ $ and $\ \Omega. \ $

{\bf 5.} Specially developed graph presentation of the solution set
of (\ref{res}) allows to construct isomorphism between independent
cluster of resonantly interacting waves and corresponding dynamical
systems. This yields constructive algorithm for generating dynamical
systems symbolically, for instance using MATHEMATICA.

{\bf 6.} The same approach (the use of the algorithms from
\cite{comp_all}, low boundary computing, graph construction and
generation of dynamical systems) can be used directly for any
mesoscopic system with resonances of the form (\ref{res}) or, more
generally, of the form \vskip -0.6cm \be \label{res_general} p_1\o_1
\pm p_2\o_2 \pm ... \pm p_s\o_s=0, \quad p_1\vec{k}_1 \pm
p_2\vec{k}_2 \pm .... \pm p_s\vec{k}_s=0 \nonumber \ee \vskip -0.2cm
with integer $\ p_i,\ $ in this case global low boundary will depend
on $\ \max_i |p_i|$.

{\bf Acknowledgements.} Author acknowledges the support of the
Austrian Science Foundation (FWF) under projects SFB F013/F1304.
Author expresses special gratitude to S. Nazarenko and M.Tanaka for
stimulating discussions; to Ch. Feurer, G. Mayrhofer,
 C. Raab and O. Rudenko for MATHEMATICA programming, and to anonymous
 referees for valuable recommendations.

\end{document}